\documentclass[%
  preprint,
 showpacs,
 showkeys,
 preprintnumbers,
 amsmath,amssymb,
 aps,
  pra,
  longbibliography,
 floatfix,
 ]{revtex4-1}

\usepackage{mathptmx}% http://ctan.org/pkg/mathptmx

\usepackage{amssymb,amsthm,amsmath}

\usepackage{tikz}
\usepackage[breaklinks=true,colorlinks=true,anchorcolor=blue,citecolor=blue,filecolor=blue,menucolor=blue,pagecolor=blue,urlcolor=blue,linkcolor=blue]{hyperref}
\usepackage{graphicx}% Include figure files
\usepackage{url}

\usepackage{xcolor}
\usepackage{tikz}
\usetikzlibrary{decorations.pathreplacing,decorations.markings}

\begin{document}

\title{Towards fractal gravity}

\author{Karl Svozil}
\email{svozil@tuwien.ac.at}
\homepage{http://tph.tuwien.ac.at/~svozil}

\affiliation{Institute for Theoretical Physics,
Vienna  University of Technology,
Wiedner Hauptstrasse 8-10/136,
1040 Vienna,  Austria}
%
%\affiliation{Department of Computer Science, University of Auckland,
%Private Bag 92019, Auckland, New Zealand}

\date{\today}

\begin{abstract}
In an extension of speculations that physical space-time is a fractal which might itself be embedded in a high-dimensional continuum, it is hypothesized to ``compensate'' for local variations of the fractal dimension by instead varying the metric in such a way that the intrinsic (as seen from an embedded observer) dimensionality remains an integer. Thereby, an extrinsic fractal continuum is intrinsically perceived as a classical continuum. Conversely, it is suggested that any variation of the metric from its Euclidean (or Minkowskian) form can be ``shifted'' to nontrivial fractal topology. Thereby ``holes'' or ``gaps'' in spacetime could give rise to (increased) curvature.
\end{abstract}

%\pacs{03.65.Aa, 03.65.Ta, 03.65.Ud, 03.67.-a}
\keywords{entanglement, quantum state, quantum indeterminism, quantum randomness}
%\preprint{CDMTCS preprint nr. x}

\maketitle

Embedded observers and agents~\cite{bos1,toffoli:79,svozil-94,roessler-98} are
operationally bound by
self-reflexive, intrinsic methods and means available from within the very system they exist.
Such observers have no access to extrinsic, Platonistic entities which are beyond their operational physical capacities.
(They may, nonetheless, have inspirational ``afflatus'' or ideas about some external truth; but would not be
able to prove this in any effable way~\cite{Jonas-ineffability} beyond zero-knowledge proof methods.)
Indeed the situation embedded observers have to cope with appears much more severe as in the
{\em allegory of the cave} mentioned by Plato~\cite[Book~7, 514a-517e, p.~220-223]{plato-republic},
in that the latter assumes the existence of a supposedly ontologic level: an observer can be {\em ``dragged right out into the sunlight.''}
The assumption of such ontologic level could, from an idealistic stance~\cite{stace}, be considered problematic,
as any observer appears to be permanently captivated in a Cartesian prison~\cite[Second Meditation, 26-29, p.~17-20]{descartes-meditation}
(see also Putnam's ``brain in the vat'' metaphor~\cite[Chapter~1]{putnam:81}, among others),
and {\em ``in the strict sense only a thing that thinks.''} Idealistic philosophy has it that~\cite{Goldschmidt2017-idealism-Ch3},
{\em ``the world is mental through-and-through.''}
Poincar\'e has pointed out in the introduction to {\em La valeur the la science}~\cite{poinc06,Poincare-Halsed-2013},
{\em
``Does the harmony the human intelligence thinks it discovers
in nature exist outside of this intelligence?
No, beyond doubt a reality completely independent of the mind which conceives it, sees or feels it, is an impossibility.
A world as exterior as that, even if it existed, would for us be forever inaccessible.''}

Therefore, when it comes to the formalization of physical theories, any such framework
ought to include and use, as much as can be possibly afforded, intrinsic, that is,  operationally feasible, elements of physical description~\cite{bridgman}.
Gaussian geometry, for example, characterizes a surface with totally intrinsic methods~\cite[Section~3.2, p.~46,47]{1993-Nottale}.
It appears prudent to include {\em epistemic} considerations rather than uncritically assume that one deals with ontic elements of perception.
Poincar\'e's and even more and explicitly so Einstein's conceptions and constructions of space and time
follow this pursuit in that they operationalize physical time by conventionalizing,
in particular, time synchronizations.

Nevertheless, quasi-extrinsic perspectives may shed new light on old physical subjects and concepts.
Thereby, such extrinsic formalizations and situations, suggesting and utilizing means and methods available from a hypothetical outside, external viewpoint,
may appear very different, even exotic and counterintuitive, from the point of view of embedded, intrinsic observers.
In particular,
based on Hausdorff measures and fractal dimension theory~\cite{Hurewicz-Wallman-1948,Rogers-1970,falconer2,mattila_1995,Montiel-1996,adda-2007,Edgar-08,Porchon-2012}
of fractals~\cite{mandelbrot-83},
it has been suggested that, while (i) extrinsically and ontologically,
space-time might be a fractal set with possibly non-integer dimension~\cite{Ord-83,Nottale-Schneider-84},
(ii) intrinsically and epistemically, that is, from an operational point of view,
it might appear as if observers  embedded in such fractals would experience not much phenomenological
differences as compared to ``inhabiting'' standard continua such as $\mathbb{R}^n$~\cite{,sv1,sv2,sv3,sv4,sv5}.
In other words, the fractal space-time concept can be put to some extreme by speculating
that, for all practical purposes, intrinsically embedded observers cannot differentiate between, say,
three-dimensional continua $\mathbb{R}^3$ and some continuous fractal which is a (possibly stochastic)
generalization of the Cantor set of fractal dimension three~\cite{sv4}, and which is
embedded in a larger-dimensional continuum, say, $\mathbb{R}^{d}$, with $d>3$.

I suggest here to take a further speculative step
by {\em shifting} the nontrivial topological structure of such fractals
to the metric of the (embedding) space.
Because even for non-integer dimensions,
intrinsic observers might, for all practical purposes, not
be able to differentiate
between two operationally indistinguishable premises:
they may either exist in a space
with standard (Euclidean, Minkowski) metric whose support is a fractal continuum;
or
they may inhabit a space-time whose support is a classical, integer dimensional continuum (say, $\mathbb{R}^3$),
but the Riemannian metric of the space is somehow non-standard and, in particular, non-Euclidean or non-Minkowskian.

\begin{figure}
\begin{center}
\makeatletter
\def\tikz@@@carcfinal#1#2#3{%
  \pgf@process{#2}%
  \advance\pgf@x by \tikz@lastx
  \advance\pgf@y by \tikz@lasty
  \pgfpathmoveto{\pgfqpoint{\the\pgf@x}{\the\pgf@y}}%
  #1%
  \pgfpathmoveto{\pgfqpoint{\the\tikz@lastx}{\the\tikz@lasty}}%
  \let\tikz@@@arcfinal=\carc@orig@@@arcfinal
}

\def\tikz@carcfinal{%
  \tikz@lastxsaved=\tikz@lastx%
  \tikz@lastysaved=\tikz@lasty%
  \let\tikz@arcfinal=\carc@orig@arcfinal
  \tikz@scan@next@command%
}

\let\carc@orig@@@arcfinal=\tikz@@@arcfinal
\let\carc@orig@arcfinal=\tikz@arcfinal

\def\tikz@cchar{% If keyword starts with "c..."
    \pgfutil@ifnextchar i %... starts with "ci..."
        {\tikz@circle}%
        {\pgfutil@ifnextchar h% ... starts with "ch..."
            {\tikz@children}
            {\pgfutil@ifnextchar a % ... starts with "ca..."
                {\carc@call}
                \tikz@cochar
            }
        }
}%

\def\carc@call{\tikzset{centred arc}\tikz@scan@next@command}

\tikzset{
  centred arc/.code={%
    \let\tikz@@@arcfinal=\tikz@@@carcfinal
    \let\tikz@arcfinal=\tikz@carcfinal
  },
}
\makeatother
\begin{tabular}{cccccccc}
\begin{tikzpicture} [scale=1.6]
\tikzstyle{every path}=[line width=2pt]
\draw [red] (0,0) -- (1,0);
\draw [blue] (0,0) carc[start angle=0,end angle=360,radius=1];
\node (a) at (0.5,0.20) {1};
\end{tikzpicture}
& &
\begin{tikzpicture} [scale=1.6]
\tikzstyle{every path}=[line width=2pt]
\draw [red] (0,0) -- (1,0);
\draw [blue]  (0,0) carc[start angle=0,end angle=15,radius=1] carc (30:45:1) carc(60:75:1) carc(90:105:1) carc(120:135:1) carc(150:165:1) carc (180:195:1) carc(210:225:1) carc(240:255:1) carc(270:285:1) carc(300:315:1) carc(330:345:1);
\node (a) at (0.5,0.20) {1};
\end{tikzpicture}
\\
(a)    & &    (b)
\\
\\
\begin{tikzpicture} [scale=1.6]
\tikzstyle{every path}=[line width=2pt]
\draw [red] (0,0) -- (1,0);
\draw [blue] (0,0) carc[start angle=0,end angle=180,radius=1];
\node (a) at (0.5,0.20) {1};
\end{tikzpicture}
& &
\begin{tikzpicture} [scale=1.6]
\tikzstyle{every path}=[line width=2pt]
\draw [red] (0,0) -- (0.5,0);
\draw [blue] (0,0) carc[start angle=0,end angle=360,radius=0.5];
\node (a) at (0.20,0.20) {$\frac{1}{2}$};
\end{tikzpicture}
\\
(d)    & &   (c)
\end{tabular}
\end{center}
\caption{
An intuitive and informal example
may help to understand
why ``cutting out holes'' in a continuum might yield different radii if one ``glues''
together the remaining pieces.
(a) consider  an original circle with radius 1;
(b) pieces of 30 degrees are cut out of (a), thereby effectively reducing the length of the set by a factor of two;
(c) those pieces are ``glued together'' to yield  a half-circle;
(d) alternatively one can draw a full circle with a reduced radius of half the original radius.
\label{2017-fg-f1}
}
\end{figure}
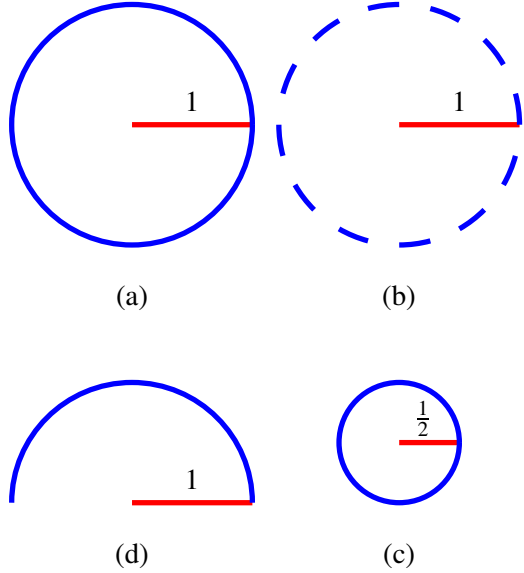

For the sake of an intuitive, informal example of why ``cutting out holes'' in a given set and ``gluing together''
the remaining pieces might affect the geometric properties of the object,
consider a situation depicted in Fig.~\ref{2017-fg-f1},
in which segments of a unit circle are eliminated, and the remaining pieces form a new circle of smaller radius.

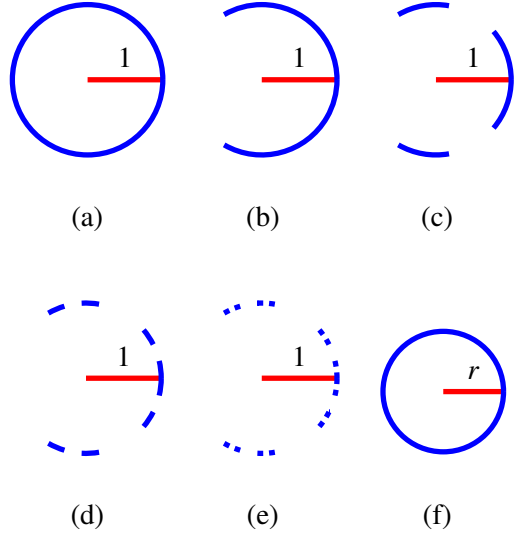
\begin{figure}
\begin{center}
\makeatletter
\def\tikz@@@carcfinal#1#2#3{%
  \pgf@process{#2}%
  \advance\pgf@x by \tikz@lastx
  \advance\pgf@y by \tikz@lasty
  \pgfpathmoveto{\pgfqpoint{\the\pgf@x}{\the\pgf@y}}%
  #1%
  \pgfpathmoveto{\pgfqpoint{\the\tikz@lastx}{\the\tikz@lasty}}%
  \let\tikz@@@arcfinal=\carc@orig@@@arcfinal
}

\def\tikz@carcfinal{%
  \tikz@lastxsaved=\tikz@lastx%
  \tikz@lastysaved=\tikz@lasty%
  \let\tikz@arcfinal=\carc@orig@arcfinal
  \tikz@scan@next@command%
}

\let\carc@orig@@@arcfinal=\tikz@@@arcfinal
\let\carc@orig@arcfinal=\tikz@arcfinal

\def\tikz@cchar{% If keyword starts with "c..."
    \pgfutil@ifnextchar i %... starts with "ci..."
        {\tikz@circle}%
        {\pgfutil@ifnextchar h% ... starts with "ch..."
            {\tikz@children}
            {\pgfutil@ifnextchar a % ... starts with "ca..."
                {\carc@call}
                \tikz@cochar
            }
        }
}%

\def\carc@call{\tikzset{centred arc}\tikz@scan@next@command}

\tikzset{
  centred arc/.code={%
    \let\tikz@@@arcfinal=\tikz@@@carcfinal
    \let\tikz@arcfinal=\tikz@carcfinal
  },
}
\makeatother
\begin{tabular}{cccccccc}
\begin{tikzpicture} [scale=1]
\tikzstyle{every path}=[line width=2pt]
\path
  (0,1) coordinate(1)
  (-1,0) coordinate(2)
  (0,1) coordinate(3)
  (0,-1) coordinate(4);

\draw [white]  (0,0) carc[start angle=0,end angle=120,radius=1] carc (240:360:1);
\draw [red] (0,0) -- (1,0);
\draw [blue] (0,0) carc[start angle=0,end angle=360,radius=1];
\node (a) at (0.5,0.3) {1};
\end{tikzpicture}
& &
\begin{tikzpicture} [scale=1]
\tikzstyle{every path}=[line width=2pt]
\path
  (0,1) coordinate(1)
  (-1,0) coordinate(2)
  (0,1) coordinate(3)
  (0,-1) coordinate(4);

\draw [red] (0,0) -- (1,0);
\draw [blue]  (0,0) carc[start angle=0,end angle=120,radius=1] carc (240:360:1);
\node (a) at (0.5,0.3) {1};
\end{tikzpicture}
& &
\begin{tikzpicture} [scale=1]
\tikzstyle{every path}=[line width=2pt]
\path
  (0,1) coordinate(1)
  (-1,0) coordinate(2)
  (0,1) coordinate(3)
  (0,-1) coordinate(4);

\draw [white]  (0,0) carc[start angle=0,end angle=120,radius=1] carc (240:360:1);
\draw [red] (0,0) -- (1,0);
\draw [blue]  (0,0) carc[start angle=0,end angle=40,radius=1] carc (80:120:1) carc (240:280:1) carc (320:360:1);
\node (a) at (0.5,0.3) {1};
\end{tikzpicture}
\\
(a)    & &    (b)  & & (c)
\\
\\
\begin{tikzpicture} [scale=1]
\tikzstyle{every path}=[line width=2pt]
\path
  (0,1) coordinate(1)
  (-1,0) coordinate(2)
  (0,1) coordinate(3)
  (0,-1) coordinate(4);

\draw [white]  (0,0) carc[start angle=0,end angle=120,radius=1] carc (240:360:1);
\draw [red] (0,0) -- (1,0);
\draw [blue]  (0,0)
carc[start angle=0,end angle=40/3,radius=1]
carc[start angle=80/3,end angle=40,radius=1]
carc (80:(80+40/3):1)
carc ((80+80/3):120:1)
carc (240:(240+40/3):1)
carc ((240+80/3):280:1)
carc (320:(320+40/3):1)
carc ((320+80/3):360:1);
\node (a) at (0.5,0.3) {1};
\end{tikzpicture}
&  &
\begin{tikzpicture} [scale=1]
\tikzstyle{every path}=[line width=2pt]
\path
  (0,1) coordinate(1)
  (-1,0) coordinate(2)
  (0,1) coordinate(3)
  (0,-1) coordinate(4);

\draw [white]  (0,0) carc[start angle=0,end angle=120,radius=1] carc (240:360:1);
\draw [red] (0,0) -- (1,0);
\draw [blue]  (0,0)
carc[start angle=0,end angle=40/9,radius=1]
carc[start angle=80/9,end angle=40/3,radius=1]
carc[start angle=80/3,end angle=(80/3+40/9),radius=1]
carc[start angle=(80/3+80/9),end angle=40,radius=1]
carc (80:(80+40/9):1)
carc ((80+80/9):((80+80/9)+40/9):1)
carc (((80+40/9)+80/9):(80+40/3):1)
carc ((80+80/3):((80+80/3)+40/9):1)
carc ((80+80/3)+80/9)):120:1)
carc[start angle=240,end angle=(240+40/9),radius=1]
carc[start angle=(240+80/9),end angle=(240+40/3),radius=1]
carc[start angle=(240+80/3),end angle=(240+80/3+40/9),radius=1]
carc[start angle=(240+80/3+80/9),end angle=(240+40),radius=1]
carc ((240+80):(240+80+40/9):1)
carc ((240+80+80/9):((240+80+80/9)+40/9):1)
carc (((240+80+40/9)+80/9):(240+80+40/3):1)
carc ((240+80+80/3):((240+80+80/3)+40/9):1)
carc ((240+80+80/3)+80/9)):360:1) ;
\node (a) at (0.5,0.3) {1};
\end{tikzpicture}
&  &
\begin{tikzpicture} [scale=1]
\tikzstyle{every path}=[line width=2pt]
\path
  (0,1) coordinate(1)
  (-1,0) coordinate(2)
  (0,1) coordinate(3)
  (0,-1) coordinate(4);

\draw [red] (0,0) -- (0.801421,0);
\draw [blue] (0,0) carc[start angle=0,end angle=360,radius=0.801421];
\node (a) at (0.4,0.25) {$r$};
\end{tikzpicture}
\\
(d)    & &   (e)  & & (f)
\end{tabular}
\end{center}
\caption{
A fractal example of why ``cutting out holes'' or ``creating gaps'' in a continuum in a scale invariant manner might yield different radii if
the remaining pieces are scaled by the fractal dimension and subsequently ``pasted''
together.
(a) consider  an original circle with radius 1;
(b) the middle third segment $\left. \left[ \frac{2\pi}{3}, \frac{4\pi}{3} \right) \right.$  is cut out of (a), thereby effectively reducing the length of the set by a factor of $\frac{1}{3}$;
(c)  the middle third segments
$\left. \left[ \frac{2\pi}{9}, \frac{4\pi}{9} \right) \right.$  and
$\left. \left[ \frac{14\pi}{9}, \frac{16\pi}{9} \right) \right.$
are cut out of the remaining segments in (b),
thereby effectively reducing the length of the set by a factor of $\frac{1}{3}$;
(d)--(e) shows the iteration of this construction;
(f) alternatively one can draw a full circle with a pasting of the upscaled segments and a reduced radius $r\approx 0.8$
from Eq.~(\ref{2017-fg-shiftdrad}).
\label{2017-fg-f2}
}
\end{figure}

Another fractal example is (as often) of the Cantor set type~\cite[Section~4.10]{mattila_1995}.
Suppose from a unit circle the middle third segment $\left. \left[ \frac{2\pi}{3}, \frac{4\pi}{3} \right) \right.$
is cut out, such that the two pieces
$\left. \left[0, \frac{2\pi}{3} \right)\right. $ and
$\left. \left[\frac{4\pi}{3}, 2\pi \right)\right. $
remain, as is depicted in Fig.~\ref{2017-fg-f2}(a).
From these remaining pieces, the respective middle third segments are cut out again,
as is depicted in Fig.~\ref{2017-fg-f2}(b)--(e);
and so on {\it ad infinitum}.
Thereby a continuum of measure zero is obtained: at the $n$'th construction stage,
encode each first remaining third by 0 and each third remaining third by 1, and associate these respective
bits with the $n$'th digits of a binary number. In the limit this construction creates the binary unit
continuum $\left[ 0,1 \right]$.
However, at each construction stage, the set ``loses'' one third of its length,
so that, in the limit this length converges to zero;
that is,
$\lim_{n\rightarrow \infty} \left(\frac{2}{3}\right)^n =0$.
To avoid the scale dependence of the measure, Hausdorff introduced a non-integer exponential dimensional scale factor $d$ applied to the measure of the remaining pieces.
This ``dimension'' $d$ is defined by
an ``{\em Umklapp property}''
$d = \inf
\left\{
d\ge 0
\middle|
\lim_{n\rightarrow \infty} \left[ 2\left(\frac{1}{3}\right)^d\right]^n = 0
\right\}$,
yielding
$2^n\left(\frac{1}{3}\right)^{nd} = 2^{n+1}\left(\frac{1}{3}\right)^{(n+1)d}$, and finally $d= \frac{\log (2)}{\log (3)}$.

So, effectively, the ``price'' of scale independence of the measure is the non-intuitive fact that the dimension of this set is not a natural number.
In an {\it ad hoc} attempt to maintain some positive integer dimensionality of the set
one may go one step further and attempt to change the metric.
Thereby the intrinsic dimensional parameter is forced to become a natural number equal to or smaller than the dimension of the external embedding space.

For the sake of an example, note that the volume of a ball of radius $r$ in $d$-dimensional Euclidean space is
% http://dlmf.nist.gov/5.19#iii
$
V(d,r) = \left(\sqrt{\pi}r\right)^d /\Gamma (d/2 +1)
$.
Suppose further that this measure of volume
(which, strictly speaking, does not contain a dimensional parameter based upon Hausdorff's ``Umklapp property'' of the measure)
nevertheless has an analytic continuation for real $d\ge 0$. Then,
by ``shifting''  the dimensionality $d$ parameter to the ``curvature'' $r$; that is, by
\begin{equation}
V(d,1) = V(1,r),
\label{2017-fg-shiftdr}
\end{equation}
one obtains a ``radius'' $r$ associated with the Cantor set by inserting $d= {\log 2}/{\log 3}$; that is,
\begin{equation}
r
=
\frac{\pi^\frac{d}{2}}{2\Gamma\left(\frac{d}{2}+1\right)}
=
\frac{\pi^\frac{\log 2}{2\log 3}}{2\Gamma\left(\frac{\log 2}{2\log 3}+1\right)}
\approx 0.8.
\label{2017-fg-shiftdrad}
\end{equation}

% N[Pi^(Log[2]/(2*Log[3])) /(2 Gamma[1 + (Log[2]/(2*Log[3]))])]

By abduction one may infer the following general {\it desiderandum}
for the parametrization of ``volume'' as it relates to
fractal dimensionality and curvature:
\begin{equation}
V(d,R) = V(m,r).
\label{2017-fg-shiftabd}
\end{equation}
Thereby the terms
\begin{itemize}
\item[(i)] fractal dimension $d$ on the left hand side of (\ref{2017-fg-shiftabd}) refers to the dimension of the fractal object,
as seen extrinsically, whereby the object is embedded in a space of extrinsic, higher dimensionality $n$;
\item[(ii)] outer, extrinsic curvature, parametrized by the radius  $R$ on the left hand side of (\ref{2017-fg-shiftabd}), stands for the curvature of the fractal object within an embedding space;
\item[(iii)] target dimension $m$ on the right hand side of (\ref{2017-fg-shiftabd}), refers to the intrinsic dimension of the object ``forced'' to be a natural number;
thereby the fractal set will, operationally and intrinsically, not be perceived as fractal but rather as a conventional continuum
$\mathbb{R}^m$ of smaller or equal dimensionality than the embedding space, but of higher or equal dimensionality than the fractal;
that is,
\begin{equation}
d \le m \le n;
\label{2017-fg-shiftabdeod}
\end{equation}
\item[(iv)]  intrinsic curvature, parametrized by the radius $r$ on the right hand side of (\ref{2017-fg-shiftabd}), refers to the curvature experienced intrinsically upon pretension of
the  target dimensionionality.
\end{itemize}
Corresponding to (\ref{2017-fg-shiftabdeod}), as compared to the extrinsic radius, one obtains a smaller or equal
intrinsic radius; that is
\begin{equation}
R \ge r.
\label{2017-fg-shiftabdeod2}
\end{equation}

Nottale~\cite[Section~4.5]{Nottale-2011} (for earlier discussions see Refs.~\cite[Section~3.10]{1993-Nottale} and~\cite{NOTTALE20011577}) and
Nottale, C\'el\'erier, and Lehner
have suggested a different, somewhat converse, ``dual'' approach by considering a scale relativity for gauge field theories, which is based upon~\cite{Nottale-2006}
{\em ``curvature at large scale and fractality at small scales.''}
Thereby~\cite[Section~4.5.3, p.~129]{Nottale-2011}, {\em ``the metric elements and
its curvature are everywhere explicitly scale dependent and divergent when
the resolution scale tends to zero.''}
This approach has been motivated by an {\it a priori}, given, fractal support of field theory.
It presents no attempt to ``re-encode'' or ``renormalize'' the curvature and the metric
in the presence of a fractal support such that this support intrinsically appear trivial in its topology.

Of course, these considerations are tentative, highly speculative and need further scrutiny.
To quote a Referee, {\it ``the formal derivation remains an open question.''}
Many issues and questions remain, among them how to  conceptualize the shift (back \& forth)
from the ``fractality of the continuum''
to the metric; and {\it vice versa} in more general situations.
Also, it needs to be seen how to obtain curvature from an originally flat (zero curvature) spacetime.

In the end, there might appear a possibility to extend the formalism of general relativity by
``punching (possibly scale invariant) holes'' or ``gaps'' into space-time;
the remaining parts being intrinsically ``stitched together''; thereby rendering
a  theory of gravity which generalizes, or at least offers an alternative viewpoint to,
relativity theory by assuming a fractal geometric support
with non-curved standard metrics.

\begin{acknowledgments}
The idea to fractal gravity emerged from a skype conversation of the author with  Hui Deng, Tyler Hill and Barry Sanders on July 5th, 2017 on their beautiful paper~\cite{Hill-2017} on light scattering in any dimension.
Further formalizations were discussed and investigated with Ludwig Staiger and Cristian S. Calude.
I kindly thank Karin Verelst for her interest in epistemological matters, and for her suggesting the quote by Henry Poincar\'e.
I am grateful to Thomas Sommer for discussions and a critical reading of the manuscript.
Nevertheless, I am to blame for all the misconceptions and omissions.
\end{acknowledgments}

%\bibliography{svozil}


\end{document}